\documentclass[11pt]{article}
\newcommand{\be}{\begin{equation}}
\newcommand{\ee}{\end{equation}}
\usepackage{amsmath}
\usepackage{amsfonts}
\usepackage{amssymb}

\usepackage{graphicx}
\begin{document}

\begin{center}
{\Large{\textbf{Telekinetic Entanglement}}}\\ 
Johann Summhammer\\ 
{\small{Vienna University of Technology,
Institute of Atomic and Subatomic Physics\\ 
Stadionallee 2, 1020 Vienna, Austria\\}}
\end{center}

\begin{abstract} A numerical thought experiment with two momentum correlated particles is presented, in which particle A passes through a series of zig-zagging slits and particle B moves unobstructedly. It is shown that, if particle A's meandering path is monitored by successive non-detections, particle B will loosely adhere to a similar trajectory without violating momentum conservation. The discussion relates this apparent telekinetic influence, which is a standard quantum mechanical result, to supposedly real telekinesis and to Stapp's hypothesis on intention in quantum physics. \end{abstract}

\section{Introduction}

While looking for inspiring science fiction I discovered some demonstrations of how to do telekinesis on youtube. The particular videos I mean are several years old and feature an adolescent boy named Nick who exhibits a healthy disdain of esoterics and comes across very amiably. I was especially intrigued by his first lecture \cite{Nick1}. Nick demonstrates how he moves sun glasses on the table in front of him, apparently without any physical means. Then he explains how to achieve this: One should move the object by hand many times in the way one intends to do telekinetically later on, and in doing so one should memorize every aspect of it; the motion of the object, of the arm, the details of the environment, etc. Then one should take the hand off the object, concentrate, and visualize the exact motion of the object and the arm, but without actually moving the arm. And, with some luck, the object should move. I don't want to discuss the credibility of the video. Rather, I want to look at it with a quantum mechanical bias and take it as an instruction for a specific experiment. Translated into quantum physics, what Nick says is essentially this: Understand the object and its representation in memory as two systems. Establish strong entanglement between the motional degrees of freedom of the object and those of its representation in memory, whichever degrees of freedom the latter may be in the nervous system and perhaps in the body as a whole.\footnote{The formation of memory is a chemical process not yet fully explained, in which different mechanisms seem to be involved, e.g. \cite{NeuroMetals, NeuroDNA}. Since chemical reactions are mostly exothermal, the final products are very unlikely to be entangled with degrees of freedoms to which entanglement existed before the reaction. We therefore have to suggest other components in the nervous system and in the surrounding tissue, with which entanglement is possible.}
Then do successive mental measurements on this representation. Each such measurement re-distributes the relative weights of the degrees of freedom of the representation. As different distributions correspond to different visualizations, the whole series of mental measurements will appear as some specific motion of the object in your imagination. Due to entanglement, the motional degrees of freedom of the object itself will also be re-distributed with each mental measurement, and this results in changes of its position. 

Leaving aside the - presumably astronomical - veto from thermodynamics, it is clear that a \emph{controlled} influence on one system of a pair of entangled systems will have no effect whatsover on the other, because this would violate the no-signalling principle, momentum conservation, etc. However, if we allow one system to randomly follow, or not follow, a previously set out path, and it accidentally does follow that path, the entangled system will mimick a similar path, at least for some time. In the next chapters I want to illustrate this. For this purpose I will analyze the behavior of two momentum entangled particles, where one may pass through a maze of slits, while the other is completely free. 

\section{Conceptual experiment}
Figure 1 shows a sketch of the conceived experiment. A source emits two momentum entangled particles of equal mass in opposite directions. Particle A moves towards a number of walls, which are parallel to the y-axis, and which may have arbitrary distances between them. Each wall has a slit of arbitrary width and position through which particle A may pass and proceed to the next wall. If the relative positions of the slits happen to be such that a direct line of sight through them is not possible, particle A can pass through the whole series only by repeated diffraction. Each wall also functions as a detector which fires when hit by particle A. If a wall fires, the experiment ends prematurely. If it does not fire at the time we expect particle A to arrive there, we gain the information that particle A has passed through the slit of that wall. If none of the detectors fires we can conclude that particle A has passed through all the slits. On the other side of the source particle B is emitted. This particle is free. There are no restrictions or detectors anywhere near its possible path. Therefore we learn about its whereabouts at a given time only by inferences drawn from information obtained about the progress of particle A. We are interested in the path we can reconstruct for particle B if particle A manages to pass through all the slits. \footnote{The detector walls serve the purpose of obtaining information step by step. Replacing them by plain absorbers and a single detector after the last slit would give us the same final information.}

\begin{figure}[htp]
\centering
\includegraphics[scale=0.30]{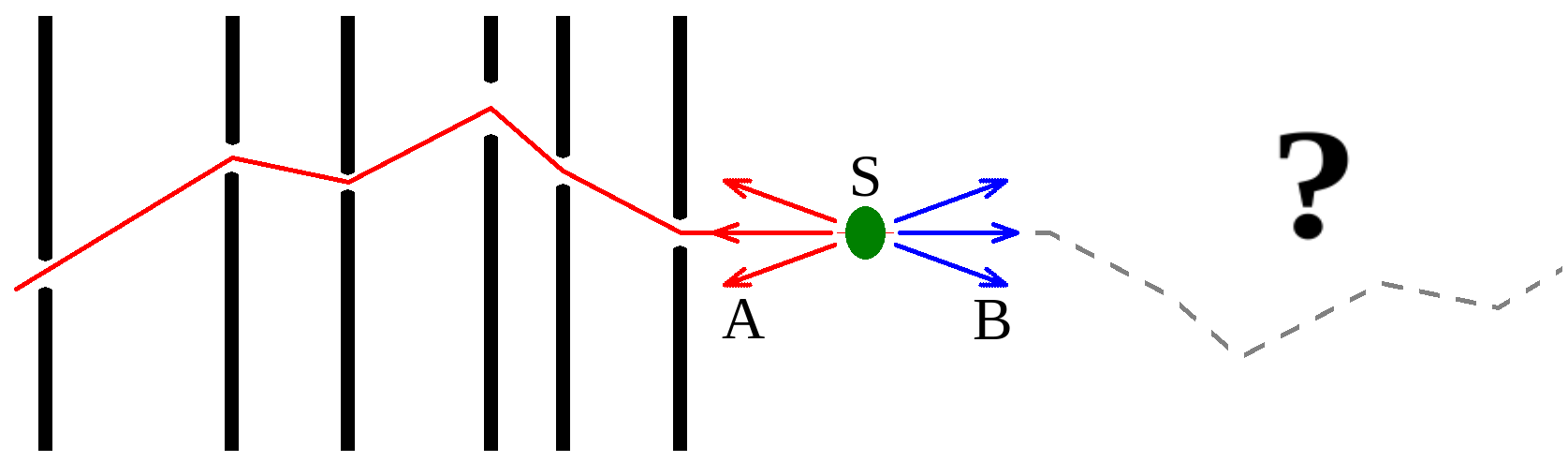}
\caption{Layout of the conceptual experiment. A source S emits a strongly momentum entangled particle pair A,B in opposite directions. Particle A can pass through a zig-zag of slits in a series of detector walls. What can we say about particle B's position while particle A successfully passes one slit after the other?}
\label{}
\end{figure}

The source emits the particle pair as a short pulse. In the horizontal direction (x-direction) the average momentum of each particle shall be very much larger than the width of its momentum spectrum in the y-direction, including the widening of the momentum spectrum of particle A during the slit diffractions. Therefore particle A's momentum along x is practically unaffected by its history along y. This allows a separation of variables for the vertical and the horizontal description and we can replace the x-position coordinate of both particles by the time since emission from the source. This, in turn, allows to reduce the description of their entanglement to a one-dimensional problem along the y-axis. The momentum and position coordinates in the following analysis thus pertain exclusively to the y-components, unless stated otherwise. The joint wavefunction of the particles can be written in the form

\be
\Psi(y_a, y_b, t) = \sum_{k_a} \sum_{k_b} \Theta(k_a,k_b) e^{ik_ay_a+ik_by_b-i(E_a+E_b)t/\hbar}.
\ee

\noindent Here, $\Theta(k_a,k_b)$ is the joint momentum amplitude distribution of the particles A and B. Throughout the analysis we use summations rather than integrals to afford a direct connection to the numerical implementation outlined in the appendix. The y-components of the energy of the particles, which are considered non-relativistic, are $E_a = {\hbar^2 k_a^2 } / {2m}$  and $E_b = \hbar^2 k_b^2 /2m$, respectively. Each time particle A successfully passes through the slit in a wall the joint momentum distribution and the associated energies change. Therefore, the time $t$ is to be understood as the time since passage of the last wall. 

The emitted pairs shall be maximally entangled with opposite momenta. If, right after the source, particle A is found with some momentum $k$, then B will be found with $-k$. The distribution of momenta shall be gaussian around $k_a=k_b=0$ with a typical momentum spread of $\kappa$. Neglecting normalization, the original joint momentum distribution thus is
\be
\begin{split}
\Theta_0(k_a,k_b) &= e^{-\frac {k_a^2}{2\kappa^2}}  \hspace{5 mm} \text{if \hspace{2 mm}} k_a=-k_b \\
  &=  0  \hspace{13 mm} \text{if \hspace{2 mm}}  k_a \ne -k_b
\end{split}
\ee
Let us now follow the evolution of the joint wavefunction. After emission at $t=0$ both particles move away from the source and at $t=t_1$ particle A reaches wall 1. If the wall detector fires, the experiment has failed and is of no further interest. If it does not, particle A has passed through the slit. The joint wavefunction right after the slit is obtained by multiplying the wavefunction just before the slit with the transmission function of slit 1, given by
\be
\begin{split}
S_1(y_a) &= 1 \hspace{5 mm} \text{if \hspace{2 mm}} y_{1, lo} < y_a < y_{1, up} \\
&= 0 \hspace{5 mm} \text{otherwise,}
\end{split}
\ee
where $y_{1, lo}$ and $y_{1, up}$ are the lower and upper boundaries of slit 1, respectively.
Therefore
\be
\begin{split}
\Psi_1(y_a,y_b,0) &= S_1(y_a)\Psi_0(y_a,y_b,t_1) \\ &= S_1(y_a)\sum_{k_a} \sum_{k_b} \Theta_0(k_a,k_b) e^{ik_ay_a+ik_by_b-i(E_a+E_b) t_1/\hbar}.
\end{split}
\ee
The joint momentum distribution after slit 1 is calculated from the spectral decomposition of that wavefunction,
\be
\Theta_1(k_a,k_b) = \sum_{y_a} \sum_{y_b}\Psi_1(y_a,y_b,0)e^{-ik_ay_a-ik_by_b},
\ee
where we have again neglected normalization. With this, the joint wavefunction for any time after particle A has passed slit 1, and before it gets to wall 2, can be written as
\be
\Psi_1(y_a,y_b,t) = \sum_{k_a} \sum_{k_b} \Theta_1(k_a,k_b) e^{ik_ay_a+ik_by_b-i(E_a+E_b)t/\hbar}.
\ee
Equations (4)-(6) represent the generic procedure of obtaining the joint wavefunction after particle A has passed through a slit from the joint wavefunction just before it gets to the slit. We can apply this procedure to any sequence of slits.

\section{Numerical example}

Figure 2 shows the results of a calculation in which particle A has to pass through the slits of 5 detector walls while particle B is free. The slits have been given different widths for the sake of generality. They are placed such that particle A cannot pass from slit n to slit n+2 without being diffracted at slit n+1, for n=1,2,3. Let us first look at the evolution of particle A. The source, as defined in eq.2, is placed at x=0 and emits with equal probability along the whole y-axis in a range of directions determined by the value of $\kappa$. Therefore, the slit in wall 1 ($W_1$) actually only serves as an initial localization of A along y.
(A source of maximally momentum entangled particles as defined in eq.2 cannot have a localization along the axis of entanglement. Localization can only be achieved at the expense of weakening the entanglement, as is done by slit 1.)
The probability distributions of the particle A before and after each slit are shown as shades of grey, where the darkest shade means highest probability. They are normalized after each slit such that the new maximum is again black. The red trajectory connects the averages of the probability distributions after each slit. As can be seen, particle A must follow a zig-zag trajectory to pass through the whole sequence.

\begin{figure}[h]
\centering
\includegraphics[scale=0.75]{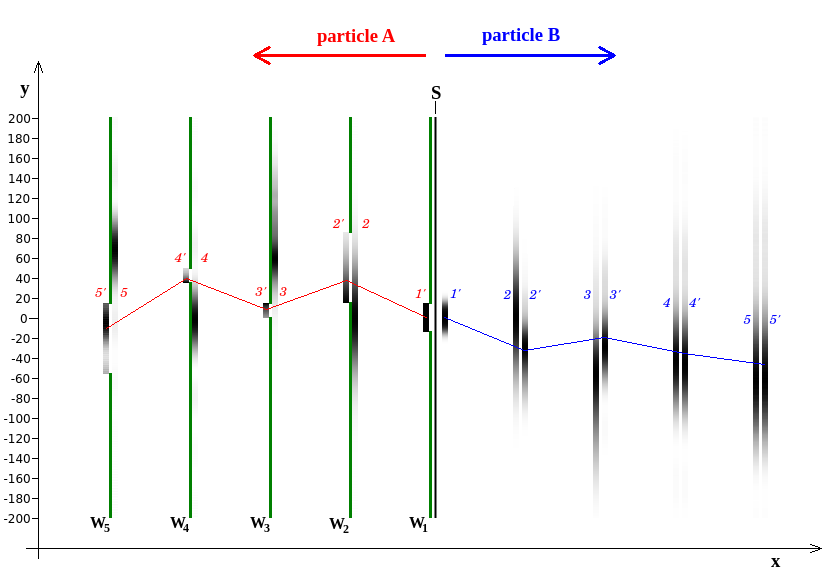}
\caption{Mean trajectories and probability distributions of particles A and B. The detector walls, through whose slits particle A may pass, are shown in green. Unprimed numbers: probability distributions at times just before particle A reaches the respective wall. Primed numbers: probability distributions right after particle A has passed the respective slit. Parameters (explained in the appendix): Width of momentum distribution $z=30$; time of flight of A from source $S$ to $W_1$: $t'=0.05$; time of flight of A between the detector walls: $t'=0.8$; upper and lower edges of slits 1 to 5: [14,-14], [85,15], [14,0], [49,35], [14,-56]. Summation limits of momentum and position: $N_k=N_y=200$. Phase space unit: $\Delta\phi=2\pi/499$. }
\label{}
\end{figure}

Now, let us look at particle B. It, too, is emitted with equal probability from anywhere along the line source into a range of directions determined by $\kappa$. In the absence of any information on particle A --- obtained as firing or non-firing at a detector wall --- we cannot localize particle B along the y-coordinate at all. But we can say something about its x-coordinate. Since we know the time of emission from the source and the average momentum along x, we can predict particle B's approximate position along x at any time. On the other hand, if we do get information on particle A, we should also be able to narrow down particle B's current y-coordinate. Let us denote by $t_i$ the time at which particle A is expected at detector wall $W_i$. If, at $t_1$, the detector $W_1$ does not fire, we learn that particle A has passed slit 1 and consequently particle B will be localized along y in an intervall essentially given by the boundaries of slit 1, but smeared out to a degree which depends on particle A's time of flight from the source to $W_1$. The two corresponding probability distributions are labelled $\emph 1'$ in Fig.2. From then on the particles evolve freely according to the new joint wavefunction. Just before A reaches $W_2$ the corresponding probability distributions for each particle have become pretty wide (labelled $\emph 2$ in Fig.2). If, at time $t_2$, detector $W_2$ does not fire, we can conclude that particle A has also passed slit 2 and we have to update the joint wavefunction. This also leads to new probability distributions, labelled $\emph 2'$ in Fig.2. By comparing particle B's distributions $\emph 2$ and $\emph 2'$ we see how the non-detection of particle A at $W_2$ instantly shifts our expectations of where we could find particle B. Now again, the particles will evolve freely, and just before particle A reaches detector wall $W_3$ their probability distributions look like those labelled $\emph 3$. If $W_3$ does not fire, they become updated 

\begin{figure}[htp]
\centering
\includegraphics[scale=0.50]{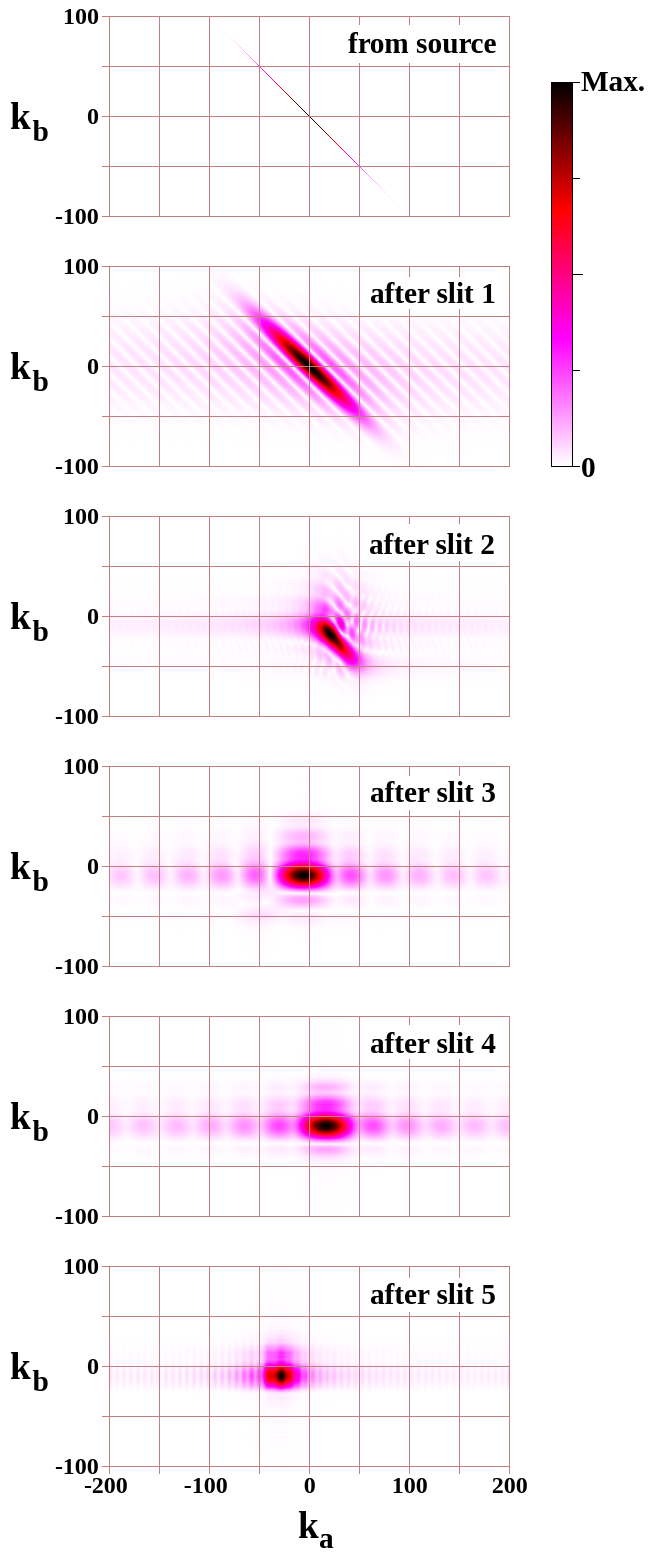}
\caption{Joint momentum distributions of particles A and B after each passage of A through a slit. The color code is linear in the modulus of the $\emph amplitude$ rather than the probability, in order to capture the wide range of diffraction effects. Renormalized after each slit so that the highest amplitude always has the darkest color.}
\label{}
\end{figure}

\noindent to those of $\emph 3'$. Here, too, one notes a strong change of particle B's expected position along y upon non-detection of A at $W_3$. And in this manner our knowledge about particle B's possible location increases with each further non-detection of particle A at a subsequent detector wall. (For the sake of completeness it should be mentioned that detection at a detector wall will of course also update our expectation about the possible location of particle B.)

Connecting the mean values of particle B's probability distributions for the times of non-detection of particle A ($\emph 1' - 5'$) gives a reconstructed mean trajectory for particle B. It is shown as a blue line in Fig.2 and has an obvious similarity to the mirror image of the mean trajectory of particle A. In particular, B's mean trajectory shows a change of direction whenever A's mean trajectory has one. Although particle B gets no physical influence from A, it looks as if the path of A were "telekinetically enforced" on B \cite{Schroedinger1935}. However, we should not fail to note that the correlation between A's and B's paths becomes weaker each time we obtain new information on A by the non-firing of a detector wall. In order to understand this, it is useful to analyze how the entanglement between A and B deteriorates with successive measurements on A. This can be done best by looking at changes of the joint momentum distribution displayed in Fig.3.

The perfect entanglement at the source appears as a diagonal line in the $k_a / k_b$-plot. Whichever momentum $k_a$ particle A may have, particle B will show the opposite momentum $k_b=-k_a$. When particle A passes through slit 1, diffraction turns each value of $k_a$ into a distribution of values. Therefore, each momentum $k_b$ will now be conjunct with a range of possible momenta of particle A. But the original entanglement "$k_b=-k_a$" is still the most dominant one, as can be seen by the central diagonal ellipse and by the fact that the state $k_a=-k_b=0$ is the most probable one. 

When particle A passes through slit 2, each of its momentum components is again diffracted into a range of momenta. This dilutes the entanglement with particle B once more. However, something of the character of the original entanglement remains, because the joint momentum distribution still shows the highest weight in a region of roughly diagonal shape. This is also emphasized by the most probable value, which belongs to a state of positive $k_a$ and of negative $k_b$. (The signs would be reversed, if slit 2 had been displaced in the opposite direction relative to slit 1.) 

\begin{table}[ht]
\caption{Evolution of particles A and B. $\bar y_a$, $\bar y_b$: mean positions right after A passed a slit; $\Delta \bar y_a$, $\Delta \bar y_b$: \emph{change} of mean positions relative to previous slit; $\Delta \bar k_a$, $\Delta \bar k_b$: \emph{change} of mean momentum relative to mean momentum before slit.}
\hspace{15mm}
\begin{tabular}{c ||rr|rr|rr}
	event & $\bar y_a$ & $\bar y_b$ & $\Delta \bar y_a$ & $\Delta \bar y_b$ & $\Delta \bar k_a$ & $\Delta \bar k_b$\\
	\hline
	A passed slit 1 & 0.00 & 0.00 & - & - & - & -\\
	A passed slit 2 & 37.85 & -33.20 & +37.85 & -33.20 & +22.68 & -19.27\\
	A passed slit 3 & 8.09 & -20.99 & -29.76 & +12.21 & -22.82 & +11.97\\
	A passed slit 4 & 39.68 & -36.36 & +31.59 & -15.37 & +20.98 & -0.67\\
	A passed slit 5 & -11.16 & -47.94 & -50.84 & -11.58 & -44.89 & +0.18\\
\end{tabular}
\end{table}

After particle A has passed slit 3 the entanglement has become very weak. The fringes from slit diffraction, which were diagonal after slits 1 and 2, are now essentially separated into vertical and horizontal stripes, as is characteristic for independent particles. The pictures are similar after particle A's passages through slits 4 and 5. Although entanglement becomes still weaker after each slit, the differences are hard to notice graphically. All one observes is that the maximum weight is either at a positive or at a negative value of $k_a$, which reflects the change of propagation direction of particle A. But for particle B it seems to remain at more or less the same negative value of $k_b$. Nevertheless, some entanglement still exists even after slit 5, and this can be seen in Table 1. The last two columns show the \emph{change} of the mean momentum of each particle which occur at the passages of particle A through slits 2, 3, 4 and 5, respectively. The values of $\Delta \bar k_a$ and especially their alternating signs are determined by the zig-zag arrangement of the relative positions of the slits. And interestingly, the same alterations of sign can be seen for the values of $\Delta \bar k_b$ (although the magnitudes quickly decrease), which proves a trace of entanglement between the particles even after particle A has passed through slit 5.

\section{Improvements}

The decrease of entanglement between the two particles whenever particle A passes through a slit naturally raises the question whether more efficient "telekinetic guidance" of particle B is possible. One option is to replace the rectangular slits by gaussian slits. Such a slit is fully transparent in the center, and becomes increasingly absorbing (=detecting, in principle) away from the center. This eliminates the diffraction sidebands in the joint momentum distribution. 

\begin{figure}[htp]
\centering
\includegraphics[scale=0.75]{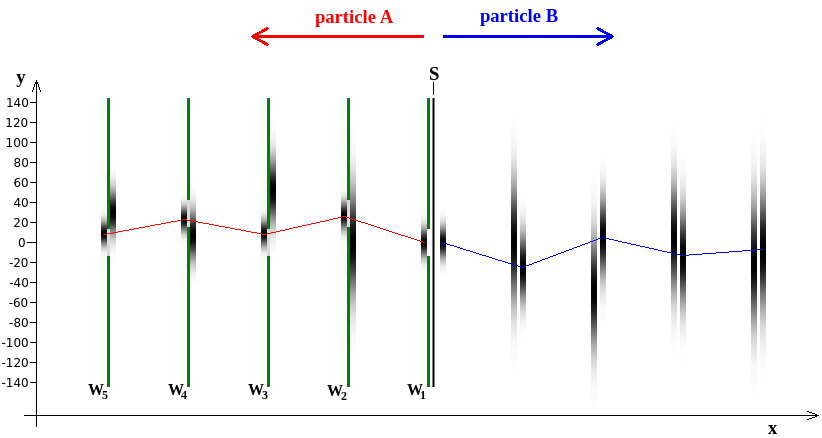}
\caption{Mean trajectories and probability distributions of particles A and B when particle A has to pass through gaussian slits. For clarity, the slits are drawn with sharp edges. The transmission probability at these edges is 1/e (37\%). Coordinates of these edges for slits 1, 3 and 5: [-14,14]; for slits 2 and 4: [15,43]. Other parameters as in the example of Fig.2.}
\label{}
\end{figure}

\noindent It still weakens the entanglement, because the passage through a gaussian slit will transform every momentum $k_a$ into a gaussian distribution of momenta. An example is shown in Fig.4. There, all slits are gaussian and have the same width. Compared to the original example (Fig.2) the mean trajectory of particle B is now noticeably a more truthful copy of the mean trajectory of particle A.

The most general option is to act directly on the momentum space of particle A. Here, too, it is important that the manipulations be non-unitary. This could mean to eliminate some momentum components or to weaken them by partial absorption (as with the gaussian slits). The reason is that a unitary operation on particle A cannot give us new empirical information on A and will therefore have no effect on particle B's momentum or position distribution, which is always conditioned on what we currently know about A. A general operation on A's momenta would have the form
\be
\Theta'(k_a,k_b) = \sum_{k'_a} M(k_a, k'_a) \Theta(k'_a,k_b)
\ee

\noindent where $M(k_a, k'_a)$ is some non-unitary matrix with complex coefficients. The difference to a unitary matrix could in principle be actualized as partial detectors, from whose non-detections the reconstruction of particle B's new whereabouts can be accomplished.   

\section{Some related experiments}

I am not aware of experimental realizations of the thought experiment discussed here. But the essential element, which is entanglement of the two particles in a high dimensional or even infinite-dimensional Hilbert space has been used in quite many experiments. All of these experiments have a different motivation and therefore are usually satisfied with detecting one particle here and the other particle there, without actively monitoring the evolution of at least one of them by a bank of detectors as in our thought experiment. Nevertheless it is worth while to comment on some of them, which could be extended toward our thought experiment. Since by now entanglement has been obtained for so may different kinds of systems, the cited references can only give a random glimpse.

Many experiments on entanglement use photon pairs from a down conversion source. Since down conversion conserves momentum, energy and angular momentum without restricting the individual photon to a specific value of these quantities \cite{Saldanha2012}, it is very close to the idealized properties of our source.

An experiment, which is already one step in our direction, was performed by Dorfer in 1998 \cite{Dorfer}, \cite{Zeilinger1999}. It exploited the entanglement of the transverse momenta of two photons. The aim there, however, was to demonstrate that, if photon A, say, is sent through a double slit arrangement, it may either show an interference pattern or just two single slit patterns, depending on whether the measurement on photon B does or does not permit to infer through which slit photon A has passed or will pass. 

In principle, one can also use the entanglement of two photons in forward momentum, or equivalently, in energy, as applied by the group of Gisin \cite{Tittel1998,Gisin2002,Gisin2007,Rogers2016}. Here, the two particles are sent into two different glass fibers. The energy of each individual photon may have a certain spread, but the total energy of the pair is well defined, thereby correlating the pair in energy. One could insert a series of fast switches along the fibre of one photon, which can be switched between the states of opaque and transparent very quickly. This would allow to cut the travelling wavepacket in a desired fashion, just as with the slits in our thought experiment. Coincidence measurements on the entangled wavepacket travelling down the other fibre will then show corresponding modifications.

Another possibility is the entanglement of two photons in the discrete Hilbert space of orbital angular momentum states. In these experiments the circular cross sections of the two beams are modulated in amplitude and phase, so that each beam is described as a superposition of the discrete orthonormal set of radial and angular momentum states, and under the right experimental conditions a high degree of entanglement can be achieved \cite{Mair2001, Vaziri2002}. The walls with slits of our thought experiment could be replaced by circular disks with appropriate radial slits placed in the beam of one photon.

An interesting option is the entanglement of two systems where each consists of many atoms, because it goes towards macroscopic massive bodies. In the experiments of the group of Polzik, e.g. \cite{Julsgaard2001, Krauter2011}, two distant gas cells, each containing about $10^{12}$ Cs atoms are manipulated by laser pulses to create entanglement between the collective angular momentum of one sample to that of the other. Due to the high number of particles the angular momentum is virtually a continuous quantity like in our thought experiment. The equivalent of slits would have to be implemented by proper timing of laser pulses applied to one sample only.

It would also be possible to use two ions, each suspended in a harmonic trap \cite{Blatt2011,Wineland2011}. The two traps would at first have to be near enough to allow interaction and thus entanglement formation between the ions. By appropriate initialisation the original state could be a superposition of a large number of oscillator quanta. Then the two traps could be moved from each other adiabatically, so as not to loose the entanglement. Successive rough measurements of position on one of the ions, e.g. by non-detection of fluorescence after stimulation with a laser beam patterned in a form analogous to a wall with a slit,  should result in a corresponding localisation of the other, very similar to our thought experiment.

\section{Discussion}

The changing directions of particle B's mean path in figures 2 and 4, and the associated changes of mean momentum as shown in the last column of table 1, immediately lead to the question how this behavior conforms to momentum conservation. Here we have to recall that particle B's probability distributions are conditional, and that the conditions are added up for each further passage of particle A through a slit: \emph{Given} that particle A has passed through slit 2, particle B's y-coordinate will be found to adhere to the distribution \emph{2'}, and later to \emph{3} (and to a continuum of distributions for the times between); and \emph{given} that particle A has passed slits 2 \emph{and} 3, particle B's y-coordinate will be found to adhere to the distribution \emph{3'}, etc. This implies that, when particle A passes through a slit, the amplitude of any of the momentum eigenstates of particle B, whose superposition localizes particle B, can only become smaller or at best remain the same. A change of the average momentum of B thus happens by a relatively stronger decrease of those momentum components, which determined the former propagation direction, and thereby allowing other momentum components to gain in relative weight and so determine the new propagation direction and corresponding localization of particle B.

We can look at this question also from the point of view of the count statistics of a real experiment, which would be performed with many particle pairs A and B. If we measure the mean momentum of particles A after a certain slit, we will get a specific value. If we then measure the mean momentum after the next slit, we will get another value, because certain particles were absorbed by the wall and will no longer be counted. And the same happens for the particles B, because we always count only those whose partner A passed the slit. The change in momentum between the two ensembles of particles A before and after a slit is thus taken up by the wall. And for particles B the difference lies in those particles B which are no longer counted, and left flying along, because their partner A was absorbed by the wall.

Let us now return to Nick's video. It has inspired the present thought experiment, and yet we see that we have trailed off pretty far from what one could call real telekinesis. There are at least three important differences. By addressing them one by one I will argue that Nick need not necessarily have been cheating, and that real telekinesis is a quantum mechanical possibility. The three differences are:
\begin{itemize}
\item{Mental measurement versus quantum mechanical measurement with instruments}
\item{Observation of the telekinetically moving object}
\item{Intentional control versus probabilistic results}
\end{itemize}

\noindent
{\emph{\underline{Mental Measurement versus quantum mechanical measurement with instruments}}

In an ordinary quantum mechanical measurement an observer reads from an instrument a finite random integer, e.g. 1 or 0, if a wall detector did or did not fire. By mental measurement I mean an introspection in which a person reads off from his/her memory or obtains information by some other inner signal.\footnote{We do not need to go into the discussion of the Heisenberg cut here, because we take for granted that measurement results come into existence.} In the concrete case this is a measurement of the representation of the object in the human nervous system or in the body as a whole.
For this to make sense we must assume three things: the representation must exist in the form of a long lived quantum state in the body, this quantum state must be entangled with the quantum state of the object whose essential components must also be long lived, and by introspection a person must be able to make a quantum measurement on the representation. The latter is the actual mental measurement.

It is certainly the case that during the interaction of Nick with the object (this interaction consists in touching, seeing and obtaining information with any of the other senses, consciously and subconsciously, but in principle also in exchange of thermal radiation) entanglement is created between many degrees of freedom of the object and many degrees of freedom in the human body. But that this entanglement should be long-lived is practically impossible at typical biological temperatures, especially if it involves motional degress of freedom of atomic nuclei, because coherence times will not go much beyond a few cycles of thermal vibrations of typical atoms, which means picoseconds. The formation of memory as a chemically concluded process might allow somewhat longer coherence times, if we link it to the emission of photons, but with a few nanoseconds they are still very short \cite{BioCoherenceTime}. We will therefore have to think of collective degrees of freedom of macroscopic amounts of atoms, e.g. low frequency phonons or other decompositions of center of mass motion. Perhaps it is the collective electronic degrees of freedom which are important, rather than those of the massive nuclei, as might be supported by Bose Einstein condensation at room temparature in plastic \cite{PlasticBEC}. At the side of the object the problem of coherence time poses itself in the same manner, although here it is easier to imagine that the repeated motion of the object by hand, which according to Nick is necessary, should mainly couple to collective degrees of freedom. And then we still have to demand that introspection, or the attempt to imagine the object, performs a quantum measurement on the representation of the object in the human body. Conceptually we can separate this into mentally setting what is to be measured, e.g. that one wants to imagine the object at a certain place, and then seeing whether this really pops up in the imagination. The first step corresponds to defining the measurement apparatus. The second step is the "wait and see" of any quantum measurement, because the outcome is probabilistic. Altogether, the required assumptions are extreme from any serious physics point of view.

Nevertheless, let us assume, for the sake of argument, that they are fulfilled, and describe what Nick is doing. First, Nick imagines the object to be set in motion. In quantum mechanical terms this means that he makes a mental measurement of momentum on the representation. And with the quantum mechanically defined probability he may obtain the result in which his imagination shows the object in motion away from him. This actualizes certain ranges of the momentum determining parameters of the representation and, due to entanglement, it also actualizes certain ranges of the momentum determining parameters of the real object. (A quantum mechanical object at rest is always in a balanced superposition of motion in all possible directions.) Consequently, the real object will also be found in motion. 

If Nick wants to change the direction or the velocity of the motion of the object, he must do another mental measurement which probes the parameter space of the representation for the new momentum. Again the desired range of parameters may come true with the quantum mechanically defined probability and appear before his imagination as object in the new motion, provided that the prior mental measurement did not reduce to zero some of the amplitudes of the parameters necessary for the representation of the new momentum. But if his imagination does show the object with the new momentum, entanglement ensures that the real object will also be actualized in the new motional state.
\vspace{2mm}

\noindent
{\emph{\underline{Observation of the telekinetically moving object}}

In the thought experiment with particles A and B above no measurements were done on B. If such measurements had been done, this would have changed particle A's position distribution and would have weakened the entanglement, just as it occurred the other way around. The same must happen to the entanglement between Nick's sunglasses and their representation in Nick's organism, if the sunglasses are observed in motion, as was actually done by the registration on video and by the multitude of factual observations by the thermal environment. At the same time, new entanglement could be created if Nick observed the motion. As this will involve much less interaction than there is interaction with the environement, the overall effect should still be deterioration of entanglement. From this we can conclude, that Nick should find it difficult to move the sunglasses telekinetically along a complicated path, or to exercise control for more than a few steps. Interestingly, this is what he admits.
\vspace{2mm}

\noindent
{\emph{\underline{Intentional control versus probabilistic results}}

When watching Nick's video one gets the impression that his telekinesis, although requiring a specific mental effort, works pretty reliably. This is surprising, because from a quantum mechanical perspective events occur with probability only, and one would expect telekinesis to succeed only very rarely, to say the least. For instance, in our examples with particles A and B above, particle A, once it has passed slit 1, makes it through the four remaining slits only with a probability of around $10^{-4}$. For macroscopic bodies with a tremendously larger Hilbert space, this number would be smaller by many orders of magnitude. Therefore, one may wonder, whether the mental effort of intending or wishing something to happen, does have an influence. In quantum theory, but also in classical physical theories, this is not possible, because none of them has any place for human intervention or the human ability to choose. All physical theories allow subjective states at best a parallel existence to the deterministic or probabilistic evolution of physical law.

Nevertheless, in the natural sciences we take for granted the role of human intention as a faculty to be made use of. We do not doubt that an experimenter can choose to design an experiment in this or in that way. Moreover, choices are an integral part of giving meaning to tests of Bell's inequality. Many related experiments and a fair amount of physical literature would loose their power for scientific corrobaration or falsification, if we did not take choices seriously \cite{Peres}. In this regard Stapp has introduced a bold hypothesis. He suggested that human intention should not simply be put aside as a superfluous by-product in the subjective realm of an otherwise ordinary evolution of a physical system which happens to be a human organism, but should be considered to have an actual influence in the evolution of physical systems by tilting quantum mechanical probabilities in the direction of the intended results \cite{Stapp2011}. Such a tilt of probabilities must of course conform to physical conservation laws. This should be especially easy to fulfill if the tilt consists in favoring certain sub-states in a superposition of quantum states over other sub-states, because each sub-state automatically conforms to the conservation laws. If we assume Stapp's hypothesis to be correct then Nick could, by strongly willing,\footnote{Athough Peres \cite{Peres} used the term "free will" in his remark, the unthwarthed subjective experience of willing or intending seems to be the relevant aspect, and this seems to preclude choices which are experienced as unfree.} really control the outcomes of his mental measurements and telekinesis would not be the extremely rare and accidental macroscopic quantum phenomenon, but could become a useful technology.

\section{Summary and Conclusion}

I have shown a thought experiment with two momentum entangled particles, which was inspired by a video on supposedly telekinetic pushing of sunglasses on a table by a young boy called Nick. One of the two entangled particles flies towards a series of walls which act as detectors, but each wall has a slit, and the sequence of slits can only be passed if the particle follows a zig-zag path. The other particle moves freely without any restrictions. If the first particle makes it through all the slits, the other particle is found to follow a mean trajectory, which resembles that of the first. This is a straightforward quantum mechanical result. Then I tried to transfer this phenomenon to the supposed telekinesis of Nick, by postulating sufficiently long lived entanglement between Nick's organism and the object to be moved. The measurements on the first particle were replaced by introspections of Nick of how he wished and imagined the object to move, as a form of internal measurement on the representation of the object in his organism. As a consequence each such introspection should actualize new weights of the quantum mechanical momentum amplitudes of the object and thereby change its motional state. Real telekinesis would thus not violate momentum conservation and become a physical possibility. But because of the huge number of dimensions of the involved Hilbert space and the unavoidable decoherence, it would be an extremely rare occurrence, which would be next to impossible to investigate scientifically. The chances for such events could be much improved, if Stapp's hypothesis of an influence of human intention on quantum mechanical evolution, or on the statistics of quantum measurements, as I read it here, is assumed to be correct. Given the many conceivable applications of telekinesis, it may well be worth to look into this question. Even if we want to view subjective experience as an irrelevant aspect of the mindless workings of nature, the subjective experience of willing or intending may be an indicator or a precursor of a very peculiar form of quantum physical evolution in biological matter, which is able to bring about effects we would never have thought of otherwise. If I am to believe Nick's final video on the matter, which he entitled "The truth" \cite{Nick2} and which features a matured Nick some seven years after his first video, related investigations may be on the way in a non-public manner. But perhaps Nick only wanted to add some suspense to his story.

\begin{flushleft}

\end{flushleft}

\newpage
\section{Appendix: Numerical Calculation}

The numerical calculation proceeds by discretizing position and momentum variables and by introducing dimensionless time. 

\noindent
The sums over different position coordinates in eq.(5) make use of the replacements
\be
y_a = j \cdot \Delta y
\ee
\be
y_b = l \cdot \Delta y
\ee
where $j, l$ are integers between $-N_y$ and $+N_y$ and $\Delta y$ is the position increment, and of the replacements
\be
k_a = r \cdot \Delta k
\ee
\be
k_b = s \cdot \Delta k
\ee
where $r, s$ are integers between $-N_k$ and $+N_k$ and $\Delta k$ is the momentum increment. With the further definition of a phase space unit
\be
\Delta \phi \equiv \Delta y \cdot \Delta k
\ee
eq.(5) takes the form
\be
\Theta_1 (r,s) = \sum_{j=-N_y}^{+N_y} \sum_{l=-N_y}^{+N_y} \Psi_1 (j,l) e^{-i(rj+sl)\Delta \phi}.
\ee
For the numerical implementation of the summations of the type as in eq.(6) the term
\be
\frac{(E_a+E_b)t}{\hbar} = \frac{t \hbar}{2m} (k_a^2 + k_b^2) 
\ee
is broken down by introducing the dimensionless time
\be
t' \equiv \frac{t \hbar}{2m}(\Delta k)^2.
\ee
Then eq.(6) becomes
\be
\Psi_1 (j,l,t') = \sum_{r=-N_k}^{+N_k} \sum_{s=-N_k}^{+N_k} \Theta_1 (r,s) e^{i(rj+sl)\Delta \phi -i(r^2 + s^2)t'}.
\ee
Finally, the initial momentum distribution of eq.(2) is rewritten in dimensionless form by defining
\be
\kappa \equiv z \cdot \Delta k,
\ee
where $z$ is a positive real number, and thus becomes
\be
\begin{split}
\Theta_0(r,s) &= e^{-\frac {r^2}{2z^2}}  \hspace{5 mm} \text{if \hspace{2 mm}} r=-s \\
  &=  0.  \hspace{13 mm} \text{if \hspace{2 mm}}  r \ne -s
\end{split}
\ee

\end{document}